# Enhanced Polarizability and Tunable Diamagnetic Shifts from Charged Localized Emitters in WSe$_2$ on a Relaxor Ferroelectric

*Qiaohui Zhou[1], Fei Wang[1], Ali Soleymani[1], Kenji Watanabe[2], Takashi Taniguchi[3], Jiang Wei[1], and Xin Lu[1*]*

[1]Department of Physics and Engineering Physics,

Tulane University, New Orleans, Louisiana 70118, United States

[2]Research Center for Electronic and Optical Materials,

National Institute for Materials Science, 1-1 Namiki, Tsukuba 305-0044, Japan and

[3]Research Center for Materials Nanoarchitectonics,

National Institute for Materials Science, 1-1 Namiki, Tsukuba 305-0044, Japan.

*Xin Lu. Email: xlu5@tulane.edu.

Abstract: Strain modulation is a crucial way in engineering nanoscale materials. It is even more important for single photon emitters in layered materials, where strain can create quantum emitters and control their energies. Here we report the localized, charge-enhanced coupling between the charged localized emitters in monolayer tungsten diselenide (WSe$_2$) to the piezoelectric relaxor ferroelectric substrate. In addition to the strain effect, we observe a gigantic polarizability volume with the enhancement factor up to $10^{10}$. The enormous polarizability leads to a large Quantum-confined Stark shift under a small variation of electric field, indicating the potential of integrating layered materials with functional substrates for quantum sensing. We further demonstrate the tunable diamagnetic shift and *g*-factor with strain varying by ~0.05%, which confirms the existence of enhanced interaction between the localized oscillating dipoles and the ferroelectric domains. Our results signify the prospect of charged quantum emitters in layered materials for quantum sciences and technology.

Strain engineering provides a versatile route to modulate the properties of materials. It is particularly attractive for the two-dimensional (2D) layered materials since the atomically thin thickness offers exceptional sensitivity to external perturbations. Strain tuning is even more appealing for single photon emitters (SPEs) which are formed when free excitons are trapped by local defects, strain or moiré potentials in the 2D materials[1]. Strain has been demonstrated to be efficient in creating SPEs[2,3] and modulating their energies in monolayer $WSe_2$[4,5]. While most earlier work has focused on the charge-neutral SPEs, less has been done for the charged SPEs which can be controlled by chiral light due to the absence of exchange interaction and can work as flying qubits for the 2D spin-photon interface in quantum communication[6].

Using the inverse piezoelectric effect from a relaxor ferroelectric substrate, such as the piezoelectric $Pb(Mg_{0.33}Nb_{0.67})O_3$-$(PbTiO_3)$ (PMN-PT), is a well-established way for strain tuning, particularly at cryogenic temperatures. The PMN-PT substrate has also been used for tuning the free excitons in $WSe_2$, where the negatively-charged states exhibit the same tuning rate as the neutral exciton[7]. It is thus natural to expect a similar strain effect from the charged localized excitons (LXs). However, we discovered that the charged LXs in $WSe_2$ exhibit very different features on the PMN-PT substrate due to a localized, charge-enhanced interaction. In addition to an opposite trend of strain-tunable circular polarization compared to the free excitons, we observed a giant Quantum-confined Stark effect due to the enormous polarizability volume which also explains the larger energy shift and hysteresis compared to the neutral counterparts. Further magneto-optical spectroscopy measurements demonstrate *in-situ* strain-modulated diamagnetic shift and *g*-factor even under a small variation of strain (~0.05%). Our findings signify the potential of coupling charged SPEs in 2D materials with functional substrates or photonic devices for quantum/nano optics and quantum communication[6].

**Independent charge and strain controls**

Figure 1a depicts the schematic of a piezoelectric substrate. An external electric field parallel (anti-parallel) to the poling direction will lead to expansion (contraction) in the out-of-plane direction and compressive (tensile) strain in-plane. Compared to earlier strain-tuning in $WSe_2$[4,5,7], we added a top gate to control the doping density in $WSe_2$ and a front side pad (Figure 1b). To introduce the inverse piezoelectric effect, an external voltage (*i.e.*, strain voltage) can be applied to either the back side of the PMN-PT ($V_{BS}$) or a front pad ($V_{TS}$) close to the sample. Despite the smaller electric-field induced deformation, the unintentional doping effect from the front pad is much more insignificant (Note 1 of the Supporting Information). We note that unintended doping is a common side effect in strain engineering[7-9], and the inclusion of $V_{TS}$ is to diminish such effect. Before discussing the charged LXs, we first characterized the strain effect from our device by comparing the energy shifts of the neutral LXs to previous reports[4,5]. As shown in Figure 1c, due to the long-range electron-hole (e-h) exchange interaction, neutral LXs D1 and D2 have fine structure splitting (FSS) of ~ 0.7 meV. As $V_{BS}$ decreased, all neutral LXs blueshift with almost the same tuning rate (Figure S4), which is consistent with the poling direction (Methods). Figure 1d shows that D1 shifts by ~3.2 meV for a tuning range of 400 V (8 kV/cm), which corresponds to a strain variation of ~0.05% based on an earlier experiment[10]. Although the tunability we achieved is smaller compared to the best plot in literature[4], it is comparable to the average tuning ranges shown in previous studies[4,5]. It is important to note that by improving the interfacial quality, the transferred strain from PMN-PT to $WSe_2$ could be enhanced[11]. However, this is not the scope of our work as our focus is to demonstrate the distinct features of the charged localized states.

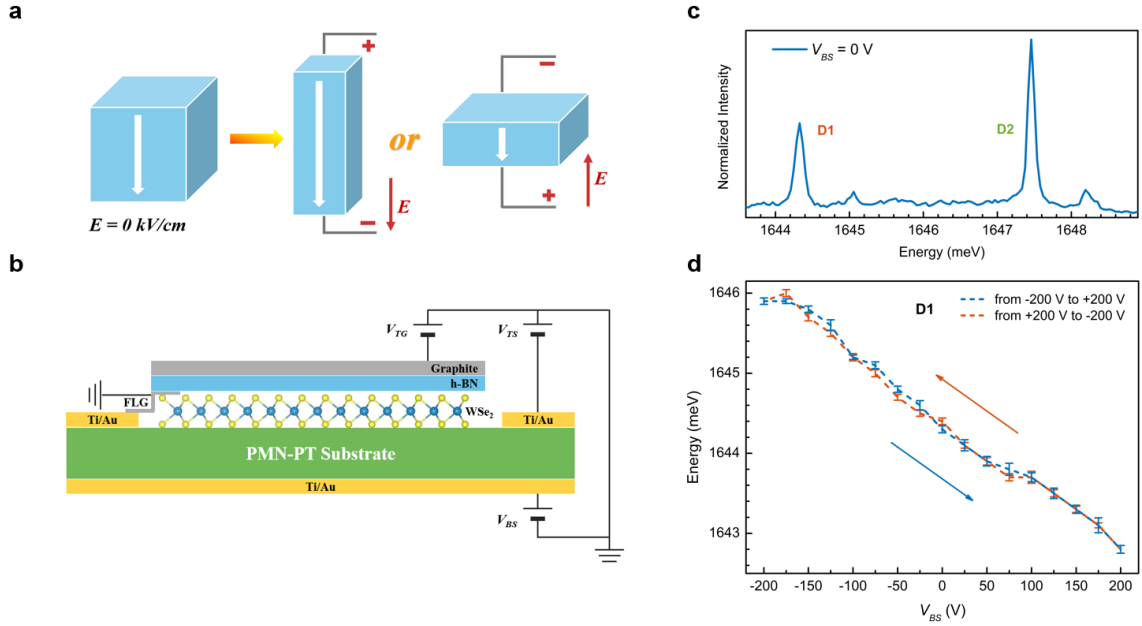

**Figure 1**. Strain-tuning of monolayer $WSe_2$ on a PMN-PT piezoelectric substrate. **a,** Sketch of a poled PMN-PT substrate with electric field applied parallel or anti-parallel to the poling direction, which produces deformations in both the out-of-plane and in-plane directions. **b,** Schematic of the device consisting of graphite, hexagonal boron nitride (h-BN), monolayer $WSe_2$ flake, and few layer graphene (FLG) on the PMN-PT substrate. Strain voltage can be applied from the back ($V_{BS}$) or through an empty metal electrode on the front ($V_{TS}$). **c,** A PL spectrum of neutral LXs, D1 and D2. **d,** Peak energies of D1 as a function of $V_{BS}$ from two sweeping directions. Error bars are from the mean deviation with 60 spectra taken at each $V_{BS}$. Excitation: 1.70 eV. Power: 150 nW.

Having demonstrated that our device is capable of strain-tuning at the cryogenic temperature, we tuned the top gate voltage ($V_{TG}$) to dope the sample with electrons. Due to the Kramers' degeneracy theorem, the signature of a singly charged LX is the appearance of a single peak with no splitting[12]. We assigned our observed single peak, S1 in Figure 2a, to be the localized singlet trion, $X_S^-$ (Note 2 of the Supporting Information). The linewidth of S1 peak is ~300 $\mu$eV, comparable to earlier studies[13,14]. While localized trions do not shift with the gate voltage when WSe$_2$ was placed on a Si/SiO$_2$ substrate[13-15], S1 redshifts from ~1646 meV to ~1641 meV with $V_{TG}$ increasing from 14 V to 18 V. Similar shift has been studied in quantum dots and explained by the quantum-confined Stark effect (QCSE)[16,17]. QCSE has also been applied to explain the shift of neutral LXs in WSe$_2$, but the required electric field is much larger; a shift of ~1 meV is achieved with an electric field of 20-100 MV/m[18]. The shift of ~5 meV that we observed is obtained when $V_{TG}$ is changed by merely 4 V, which corresponds to 26.3 MV/m after applying the Lorentz local field approximation (Note 3 of the Supporting Information). Our observation demonstrates that a shift of ~1 meV only requires for electric field of 5.3 MV/m in average. We plotted the electric field-dependent energy shift and performed fitting to extract the dipole moment ($\mu$) and polarizability volume (Figure S9). While the fitted $\mu$ = -4.7 $pm \cdot e$ = -0.23 $D$ ($1D = 3.33 \times 10^{-30}$ $C \cdot m$) is consistent with an earlier report[18], the polarizability volume is huge, reaching ~$2 \times 10^{10}$ Å$^3$. As a comparison, the polarizability volume of neutral LXs ranges from 1 to $10^3$ Å$^3$ in the h-BN encapsulated WSe$_2$ on Si/SiO$_2$[18]. We found that the $V_{TG}$-dependent shifts are independent of the strain state and exist under both compression and tension (see Figure S10 for S2 and S3 peaks). Note that the PMN-PT is poled (Method) before sample-transfer. The domains are larger after poling and oriented to some extent even when $V_{TS}$ = 0 V[19,20]. The consistently large $V_{TG}$-dependent shifts from charged LXs imply the existence of localized modulation from the ferroelectric domains in the PMN-PT substrate[21]. Interestingly, such modulation is very weak in the neutral LXs, since most of the neutral LXs do not exhibit the $V_{TG}$-dependent shift. Due to the anisotropic e-h exchange interaction, neutral LXs are linearly polarized along a specific in-plane direction[5], which could decrease the coupling strength between the localized oscillating dipoles and the ferroelectric domains.

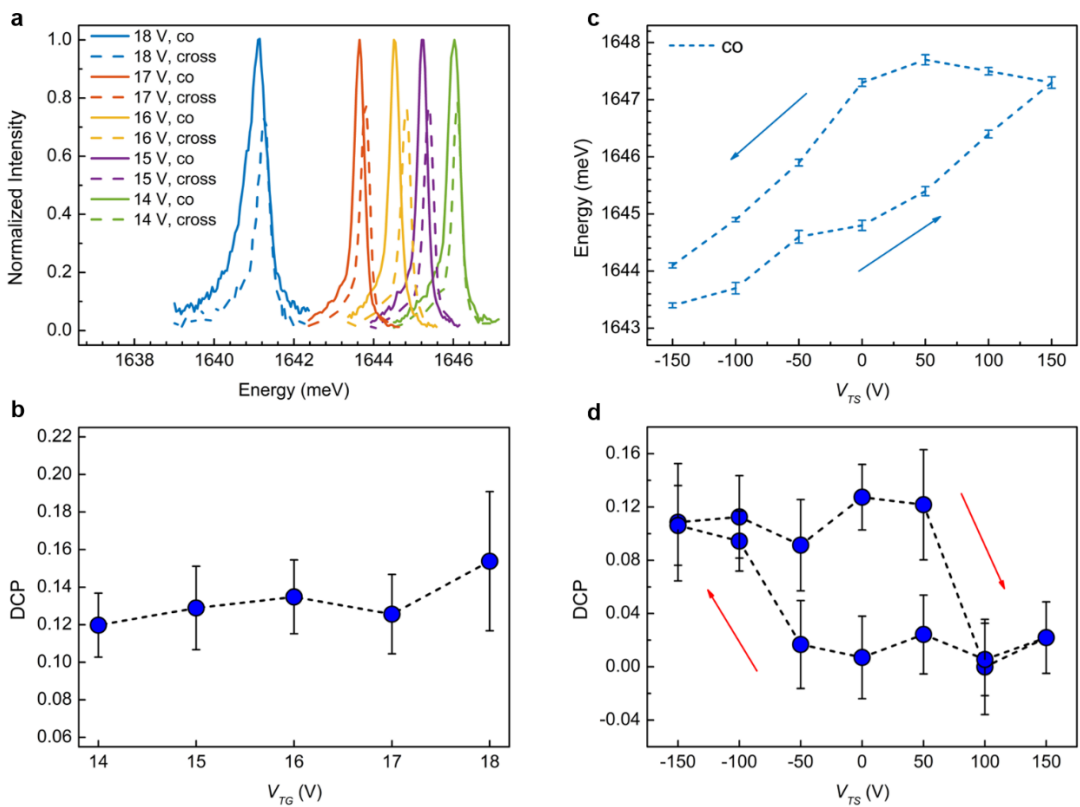

**Figure 2**. Tuning a charged localized exciton (LX) by the front strain voltage ($V_{TS}$). **a,** Circular polarization-resolved PL spectra of charged LX, S1, as $V_{TG}$ changes from +14 V to +18 V. Co-polarized (Cross-polarized) spectra are shown by the solid (dashed) lines. **b,** Degree of circular polarization (DCP) extracted from the $V_{TG}$-dependent PL measurements. **c,** Co-polarized peak energy of S1 as a function of $V_{TG}$ from two sweeping directions. **d,** DCP from two sweeping directions. Arrows in **c** & **d** indicate the tuning directions of $V_{TS}$. Error bars in **b-d** originated from the mean deviation of the peak intensities with more than 10 spectra taken for each voltage. Propagation of uncertainty leads to larger error bars in **b** & **d**. Excitation for e-h: 1.96 eV. Power: 300 nW.

We further extracted the degree of circular polarization (DCP: $(I_{co} - I_{cross})/(I_{co} + I_{cross})$) in Figure 2b and found that DCP is constant (0.14% ± 0.02%), independent of $V_{TG}$. The finite DCP of a charged LX at zero magnetic field is unexpected. The measured polarization is given by DCP = $\frac{G}{1+\frac{\tau_0}{\tau_s}}$, where $G$ is the generation rate of the co-polarized excitons, and we consider $G = 1$ in our measurements. $\tau_0$ and $\tau_s$ are the excitonic lifetime and spin-valley lifetime, respectively (Note 4 of the Supporting Information). For free excitons, a finite DCP can be observed in steady-state photoluminescence (PL) spectra because the ratio $\tau_0/\tau_s$ is small. However, $\tau_0/\tau_s$ is much larger for LXs since the lifetimes of the localized states are in the order of nanoseconds[22]. Therefore, even if the free neutral exciton exhibits co-circularly polarized emission, the neutral LXs are linearly polarized due to the exchange interaction which flips the spin-valley index within picoseconds[23]. For the same reason, the charged LX, originating from a localized trion state, is not expected to preserve any circular emission. We note that the charged/neutral LXs have been reported to exhibit finite DCP due to the proximity effect from magnetic $CrI_3$ or strained antiferromagnetic $NiP_3$ which produces an effective magnetic dipole[24,25]. However, such tiny magnetic dipole is not present in a relaxor ferroelectric. Ferroelectric gating can lead to polarization-induced doping[26,27], but it cannot boost the circularly polarized emission.

To understand the nonzero DCP, we proceeded to the strain-tuning by applying an electric field across the PMN-PT substrate. In order to mitigate the unintentional doping effect, we used the front pad ($V_{TS}$) to apply strain. $V_{TG}$ was kept at +17 V with the sample and the back side of PMN-PT being grounded during the measurements. Figure 2c shows the energy of S1 peak as a function of $V_{TS}$ in two sweeping directions. The trend of energy shift is opposite to that in Figure 1d ($V_{BS}$), since the strain voltage is applied from the front ($V_{TS}$). In other words, both positive $V_{TS}$ and negative $V_{BS}$ cause in-plane compressive (negative) strain and result in a blueshift. Compared to the neutral LXs, the charged LX shows a larger energy shift even for a smaller tuning range (See Figure 1c & Figure S4 for neutral LXs under $V_{BS}$; See Figure S2 for comparison of $V_{BS}$ and $V_{TS}$). Interestingly, we noticed that such difference between the charged and neutral states are absent for the free excitons[7]. A possible origin of the larger shift is the Stark effect resulting from the $V_{TS}$ which induces strain. Therefore, the observed $V_{TS}$-dependent shifts are results of both strain and QCSE.

The Stark shift can also explain the larger hysteresis between the two sweeps. The energy difference reaches 2.3 meV at $V_{TS}$ = 0 V in Figure 2c. This value is substantially larger than those from the neutral LXs (Figure 1c & Figure S4). Since the neutral LXs were measured simultaneously, and the difference at $V_{BS}$ = 0 V ranges from 0 to ~0.8 meV, the hysteresis cannot be a global effect from the PMN-PT substrate. The variation implies the existence of local interaction between the localized oscillating dipole and a specific ferroelectric domain which is not small with respect to localized potential[19,20]. As a result, a LX could sense the ferroelectric domain, which is right beneath the trap, resembling a local ferroelectric effect. Since ferroelectric materials exhibit large hysteresis, the charged LX thus experiences different local electric fields even when $V_{TS}$ sweeps back to 0 V. Compared to neutral LXs, the larger hysteresis from a charged LX indicates the charged states are more sensitive to the domains. This observation is also in agreement with the enhanced polarizability volume which leads to a significantly larger Stark shift in the charged LXs. Although sweeping can release some of the surface-trapped charges, these charges should affect all the excitonic states unselectively. If a charge enters the localized potential, it will dramatically change the spectral feature[14]. As we did not observe staircase-like change in spectra during the sweep, we excluded the effect from surface-trapped charges.

We further plotted the $V_{TS}$-dependent DCP in Figure 2d. The circular polarization decreases from ~10% at -150 V to ~0% at +150 V. Note that such tuning is absent in the $V_{TG}$-dependent measurements (Figure 2b), which validates the strain effect in Figure 2c even with the presence of QCSE. The same trend is reproduced from another charged LX, S2 (Figure S11) and the tuning ranges from 0% to ~25%. As DCP = $\frac{1}{1+\frac{\tau_0}{\tau_s}}$, our observed results indicate $\tau_0/\tau_s$ increases under the compressive (negative) strain. DCP = 10% (0%) implies that $\tau_0/\tau_s$ = 9 (>100). Our observation is in opposite to the trend of free excitons as shown by An et al[7]. The difference originates from the ratio of excitonic lifetime and spin-valley lifetime, $\tau_0/\tau_s$. For free excitons, as the radiative lifetime is ~1 ps, the dominant depolarization path that affects DCP is the long-range e-h exchange interaction (~ps). While the lifetimes of LXs are in the order of nanoseconds[22], it is crucial to consider slower processes which are insignificant for the free excitons. For instance, flipping of the resident electron could occur through spin-conserving intervalley scattering or spin-flip intravalley scattering. Both would effectively reduce the spin-valley lifetime of the resident electrons, which was found to range from 100 ps to tens of nanoseconds at 10 K[28]. While it cannot affect the DCP of free excitons as demonstrated by earlier experimental work[7], the valley lifetime of resident electrons would significantly modify $\tau_0/\tau_s$, hence affecting DCP. Calculations show that compressive strain reduces the spin nature of the lowest conduction band (where resident electrons are)[7,29], which implies that $\tau_s$ decreases under compression. This result is consistent with our observation that $\tau_0/\tau_s$ increases with negative strain.

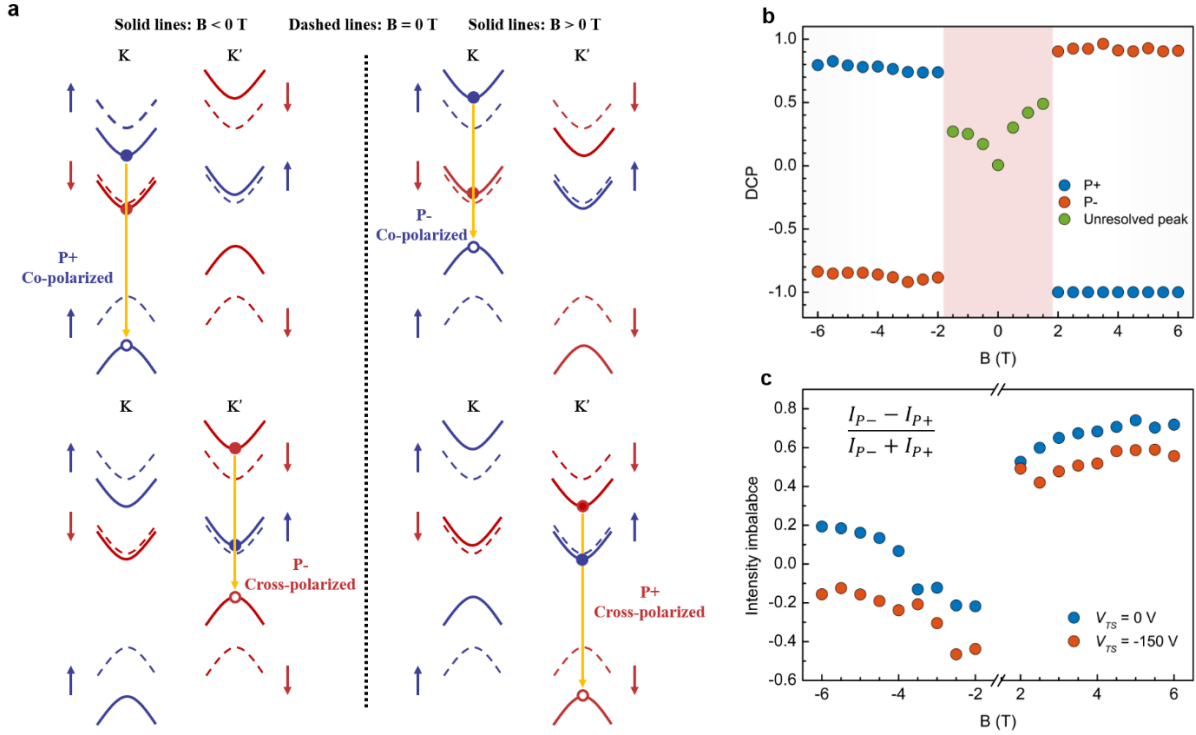

Figure 3. Circular polarization and intensity imbalance ratio of the charged localized exciton (LX) S1 under a magnetic field ($B$). **a**, Schematics of energy levels in the K and K' valleys for WSe$_2$ when $B = 0$ T (dashed lines) and $B \neq 0$ T (solid lines). **b**, Degree of circular polarization (DCP) of S1 peak at $V_{TS} = 0$ V. P + (P −) denotes the split high-energy (low-energy) peak. The two peaks are not well-resolved at low $|B|$ (pink region). **c**, $B$-dependent intensity imbalance ratio ($I_{P-}$- $I_{P+}$)/ ($I_{P-}$ + $I_{P+}$) at $V_{TS} = 0$ V, -150 V under $|B| = 2-6$ T. Excitation: 1.96 eV. Power: 300 nW

To better understand the $V_{TS}$-dependent polarization, we applied a magnetic field ($B$) in the Faraday geometry. Figure 3a shows the evolution of K and K' valleys under a finite $B$. When $B \neq 0$ T, bands in the K (K') valleys shift with different magnitudes due to the combined effect of spin, valley and orbital angular momenta[30]. Consequently, the single peak at zero $B$ would split into two. Our circular excitation is fixed and selectively couples to the K valley. As a result, excitation is in favor of the co-polarized high-energy peak (P+) for $B < 0$ T and low-energy peak (P-) for $B > 0$ T.

We first plotted the *B*-dependent DCP for S1 peak at $V_{TS}$ = 0 V in Figure 3b (See Figure S12 for DCP at $V_{TS}$ = -150 V). With zero DCP at *B* = 0 T, the co-polarized emission increases with |*B*|, regardless of the direction. As magnetic field lifts the degeneracy between the two valleys thus suppressing inter-valley scattering, the spin lifetime $\tau_s$ is prolonged and results in a higher DCP (DCP = $\frac{1}{1+\frac{\tau_0}{\tau_s}}$). Once the split peaks are well resolved (|*B*| ≥ 2 T), both P+ and P- obtain a high degree of polarization as shown in Figure 3b. It is important to note that the resonant circular excitation still excite both split peaks in our *negatively* charged LX, which is distinct from the optical initialization demonstrated in the *positively* charged LXs[13]. Due to the large splitting at the valence band maximum, $\tau_s$ is much longer for holes than electrons[31]. As a result, we were not able to selectively excite one of the split peaks even under the magnetic field. On the other hand, it also implies the more effective strain modulation to a negatively charged emitter.

While excitation is in favor of P+ (P-) for *B* < 0 T (*B* > 0 T), Boltzmann thermal distribution is to the advantage of the P- peak regardless of the direction of the magnetic field. We plotted the *B*-dependent intensity imbalance ratio, $\frac{I_{P-} - I_{P+}}{I_{P-} + I_{P+}}$ in Figure 3c for $V_{TS}$ = 0 V and -150 V when the two split peaks are well resolved (|*B*| ≥ 2 T). $\frac{I_{P-} - I_{P+}}{I_{P-} + I_{P+}}$ = 1 indicates the emission occurs after reaching thermal equilibrium. When *B* < 0 T, $\frac{I_{P-} - I_{P+}}{I_{P-} + I_{P+}}$ is around 0.2 (-0.2) at *B* = -6 T for $V_{TS}$ = 0 V (-150 V), an indication of the competing effect between thermal equilibrium and circular excitation; where thermal equilibrium favors the P- peak and circular excitation favors the P+ peak due to the helicity-dependent optical selection rule. Note that the circular excitation effect has also been discussed for the free exciton in WSe$_2$ and leads to asymmetries in the polarization-resolved emission intensities[32]. Nevertheless, the excitation effect is much less important towards *B* = 6 T with the intensity imbalance ratio approaching a constant value at ~0.7 (~0.6) for $V_{TS}$ = 0 V (-150

V). This result demonstrates that the Boltzmann distribution is the dominant effect over excitation at $B = 6$ T. We did a fitting based on the Boltzmann distribution with a factor $D = \frac{\tau_0}{\tau_0 + \tau_s}$. Qualitatively, we also obtained a decreasing $\tau_0/\tau_s$ at -150 V, consistent with the $V_{TS}$-dependent DCP in Figure 2d (Note 5 of the Supporting Information).

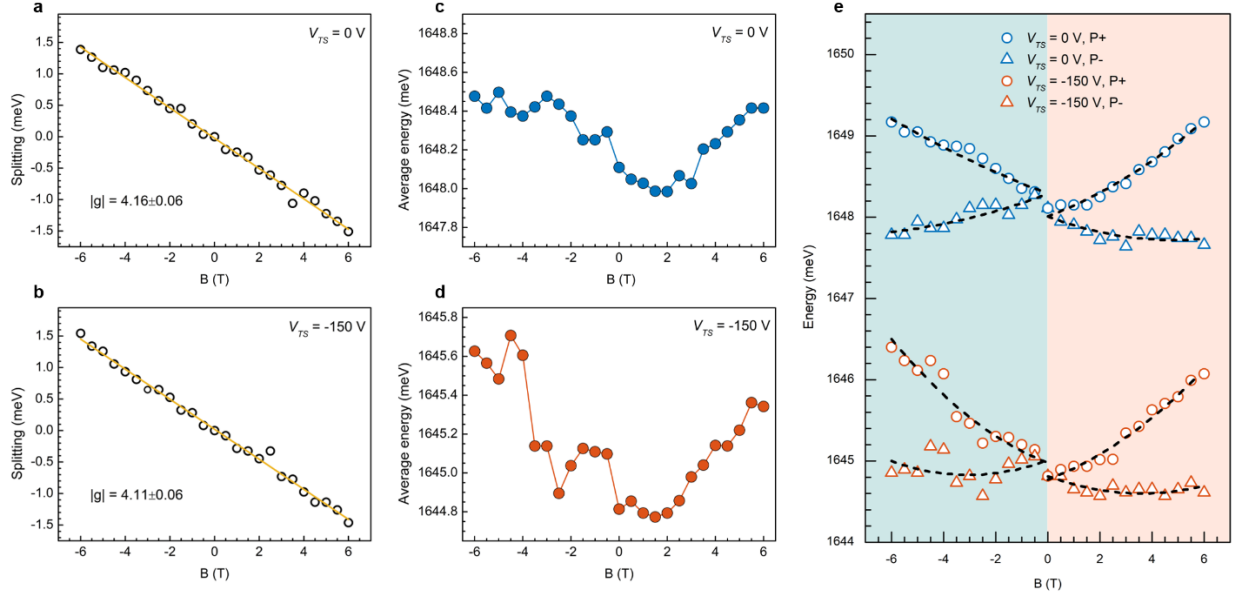

**Figure 4**. Splitting and average energies of S1 peak under a magnetic field ($B$) at $V_{TS} = 0$ V and -150 V. **a, b,** Splitting energy of S1 as a function of $B$ at $V_{TS} = 0$ V and $V_{TS} = -150$ V. **c, d,** $B$-dependent average energy, $\frac{E_{P+}+E_{P-}}{2}$, at $V_{TS} = 0$ V and $V_{TS} = -150$ V. **e,** $B$-dependent energy shifts for peaks P+ and P- at $V_{TS} = 0$ V (blue) and -150 V (red). Experimental data are shown by dots, and fitting results are displayed by the black lines. Excitation: 1.96 eV. Power: 300 nW.

We further extracted the $B$-dependent splitting ($\Delta E = E_{co} - E_{cross}$) of S1 peak and fitted the data by $\Delta E = g\mu_B B$, where $\mu_B = 58$ μeV/T is the Bohr magneton. The extracted $|g| = 4.16 \pm 0.06$ at $V_{TS} = 0$ V (Figure 4a) is consistent with that of the free trion in an h-BN-encapsulated sample[33,34]. However, it is unusual to observe $|g|\sim 4$ from the localized states in WSe$_2$ (supplementary Table 1), and the substantially smaller $|g|$ only occurs for charged LXs on PMN-PT. We also measured the g-factors at $V_{TS}$ = -150 V and +150 V (Figure 4b and Figure S14), and neither deviates significantly from 4.16. With the magnitude of g-factor around 4, we further confirmed that the S1 peak is a localized bright trion in a shallow potential trap.

We subsequently plotted the B-dependent average energy, $\frac{E_{P+}+E_{P-}}{2}$, for $V_{TS}$ = 0 V and -150 V. For Zeeman splitting of a singly charged LX, both split peaks are expected to shift linearly with the same magnitude but opposite directions. Therefore, the average energy should be a constant value within the range of spectral jittering. As shown in Figure S14, $\frac{E_{P+}+E_{P-}}{2}$ is 1653.7 ± 0.1 meV from -6 T to +6 T at $V_{TS}$ = +150 V. Nevertheless, the average energy increases by ~0.4 meV (~0.6 meV) when $|B|$ = 6 T at $V_{TS}$ = 0 V and -150 V (Figure 4c & 4d). Although diamagnetic shifts can cause the average energy increases quadratically with $|B|$, a giant field is required because of the small Bohr radius of WSe$_2$[35]. The diamagnetic shift is given by $\Delta E_{dia} = e^2 <r^2> B^2/8\mu = \sigma B^2$, where $<r^2>$ is related to the electron-hole overlap (the size of the exciton), $\mu$ is the reduced mass of the exciton, and $\sigma$ is the diamagnetic shift coefficient. Theory predicts that the diamagnetic shifts are larger for trions due to the increasing inter-particle distance. Additionally, $\Delta E_{dia}$ increases linearly with the dielectric constant of the surrounding environment[36,37]. According to the model in Ref. [36], we obtained $\sigma$ = 11 $\mu$eV/T$^2$, resulting in $\Delta E_{dia}$ ~ 0.4 meV at $|B|$ = 6 T by taking the average dielectric constants of PMN-PT[21] and h-BN, which matches very well with the shift at $V_{TS}$ = 0 V. To further confirm the diamagnetic shift and examine the tunability, we plotted the B-dependent energies of both P+ and P- at $V_{TS}$ = 0 V and -150 V (Figure 4e). The B-dependent energies of P+ and P- are given by Equations (1) and (2), respectively.

$$E(B)_{\pm} = E_0 \pm \frac{|g|}{2}\mu_B B + \frac{1}{2}\sigma B^2 \qquad (1)$$

$$E(B)_\pm = E_0 \mp \frac{|g|}{2}\mu_B B + \frac{1}{2}\sigma B^2 \tag{2}$$

where $E_0$ is the energy at $B = 0$ T. We obtained $\sigma$ = 18.8 $\mu$eV/T$^2$ for $V_{TS}$ = 0 V, and $\sigma$ = 37.8 $\mu$eV/T$^2$ for $V_{TS}$ = -150 V (Note 6 of the Supporting Information), which indicates that the ratio $<r^2>/\mu$ gets increased under tensile (positive) strain. Intuitively, tensile strain increases the distance between particles and leads to a larger value of $<r^2>$. While the reduced mass is hardly affected by strain[8], the enhanced $<r^2>$ leads to a larger $\sigma$ ($\sigma = e^2 <r^2>/8\mu$). On the other hand, $\sigma$ should decrease under the compressive strain, and no diamagnetic shift is expected, which is confirmed from Figure S14 when $V_{TS}$ = +150 V. While a small change of strain would not modify the g-factor[29], we observed a slightly larger g-factor under the negative strain ($|g|$ = 4.38, Figure S14) and considered it as an indication of stronger confinement which has been shown to increase the electron g-factor in quantum dots[38].

We note that the strain tuning of diamagnetic shift and g-factor are only present for the charged localized states (see Figure S15 for neutral LXs). While calculation predicts that a strain variation of ~1% is needed to modify the g-factor of the free bright exciton by 0.3[29], the strain modulation observed from our charged LXs only needs a change of ~0.05%. The enhanced modulation is attributed to the combined effect of localization and the presence of an extra charge: Due to the larger size of the trion, we were able to observe the diamagnetic shift even at $|B|$ = 6 T (also resulted from the larger dielectric constant from PMN-PT[21]). The tunable diamagnetic shift is practical because the localized excitonic state was in a shallow trap, which is more sensitive to an external control. As a result, a small variation of strain is capable of changing the confinement potential, thus affecting the g-factor. Meanwhile, the net charge of the charged LXs causes a stronger coupling to the relaxor ferroelectric, as evidenced by the giant QCSE. The enhanced coupling also leads to a larger hysteresis in energy from the charged LXs compared to the neutral counterparts. Due to the giant polarizability volume, it is interesting to note that charged LXs on PMN-PT could be potential quantum sensors for detecting a small electric field.

We noticed that the difference between neutral and charged states are absent in earlier experimental work on quantum dots[10], where charged states and neutral states experienced the same tuning as free excitons in WSe$_2$. The major difference in our work is that the charged LXs are in direct contact with the PMN-PT, which is impossible for quantum dots. Therefore, quantum dots can only sense the global effect from the strain tuning without seeing the localized, charge-enhanced coupling. We remark that the localized, negatively charged dark exciton (D$^-$), with both electrons at the lowest conduction bands, is very likely to have even stronger interaction with the relaxor ferroelectric. Transferring a twisted heterobilayer on the piezoelectric substrate to probe the interplay of strain and moiré potential could be another interesting topic as well[39]. As the modulation could be more effective for the bottom layer, one can achieve selective tuning of the electron or hole for the localized interlayer excitons. We note that our applied electric field is small compared to the field required to rotate the polarization axis of the charge-neutral GaAs quantum dots[40]. Further measurements with larger electric fields and magnetic fields will provide more insights into the integration of 2D layered materials with relaxor ferroelectrics and could pave way for potential applications in nanophotonics and quantum photonics.

**Methods**

**Device fabrication.** Electron beam lithography was used to deposit 5 nm Ti/100 nm Au metal contacts on both sides of the PMN-PT substrate. Before the sample-transfer process, the PMN-PT substrate was poled at room temperature by applying a direct current (DC) electric field to the backside. After reaching the targeted poling voltage (-300 V) in 5 minutes, we held the electric field for 30 minutes and released the field back to 0 V in 5 minutes.[41] Mechanically exfoliated WSe$_2$ (HQ Graphene), few-layer graphene/graphite and h-BN flakes were transferred on top of the PMN-PT substrate by a polydimethylsiloxane-based dry-transfer method.[42] The stacked device is illustrated in Figure 1b.

**Optical spectroscopy measurements**. Low-temperature magneto-optical measurements were performed in a closed-cycle cryostat (AttoDry 1000, Attocube Systems) equipped with a superconducting magnet. The samples were cooled to ~3.7 K and positioned by a piezoelectric nanopositioners (Attocube Systems). Both the 633-nm and 730-nm continuous-wave excitation lasers were collimated and focused onto the sample through an objective (NA = 0.81, Attocube Systems), with a spot diameter around 1 μm. The PL emission was collected by the same objective and directed to a high-resolution spectrometer (HRS-750, Teledyne Princeton Instruments), in which it was dispersed by a 1200 g/mm grating. A charged-coupled device (PYL-400BRX, Teledyne Princeton Instruments) was used as the detector. For polarization-resolved measurements, a λ/4 waveplate was placed after the beam-splitter, and a λ/2 waveplate, followed by a fixed-polarization analyzer, was placed before the spectrometer. Strain and top-gate voltages were applied through Keithley 6487 and Keithley 2400 source meters, respectively.

AUTHOR INFORMATION

Corresponding Author

*Xin Lu. Email: xlu5@tulane.edu.

Author Contributions

X.L. conceived the project. Q.Z. and X.L. carried out the optical spectroscopy measurements and wrote the paper, with assistance from F.W. K.W. and T.T. provided the h-BN crystals. Q.Z., F.W., and A.S. prepared the samples. X.L. and J.W. supervised the project. All authors were involved in analysis of the experimental data and contributed extensively to this work.

Notes

The authors declare no competing financial interest.

ACKNOWLEDGMENT

X.L. and Q. Z. acknowledge support from Tulane University startup fund and the Louisiana Board of Regents Support Fund (BoRSF) under award # LEQSF(2022-25)-RD-A-23. K.W. and T.T. acknowledge support from the JSPS KAKENHI (Grant Numbers 21H05233 and 23H02052) and World Premier International Research Center Initiative (WPI), MEXT, Japan. J.W. and F.W. are grateful for the support from the National Science Foundation under Grant 1752997 and the Louisiana Board of Regents under Grant 082ENH-22. We also acknowledge the support from the Micro/Nano Fabrication Facility and Coordinated Instrument Facility of Tulane University.